\documentclass[aps,prb,twocolumn,groupedaddress,amsmath,amssymb]{revtex4}

\usepackage{amssymb}
\usepackage{amsmath}
\usepackage{graphicx}
\usepackage{subfigure}
\usepackage{textcomp}
\usepackage{color}
\usepackage{amsfonts}
\usepackage{bbold}
\usepackage{dsfont}
\usepackage{epsfig}

\newcommand{\arctanh}

\begin{document}

\title{Exploring the limits of the self consistent Born approximation for inelastic electronic transport}

\author{William Lee}
\author{Stefano Sanvito}
\email{sanvitos@tcd.ie}

\affiliation{School of Physics and CRANN, Trinity College, Dublin 2, Ireland}

\date{\today}

\begin{abstract}
The non equilibrium Green function formalism is today the standard computational method for describing elastic 
transport in molecular devices. This can be extended to include inelastic scattering by the so called self-consistent 
Born approximation (SCBA), where the interaction of the electrons with the vibrations of the molecule is assumed
to be weak and it is treated perturbatively. The validity of such an assumption and therefore of the SCBA is difficult to establish
with certainty. In this work we explore the limitations of the SCBA by using a simple tight-binding model with the
electron-phonon coupling strength $\rm{\alpha}$ chosen as a free parameter. As model devices we consider Au mono-atomic 
chains and a $\rm{H_2}$ molecule sandwiched between Pt electrodes. In both cases our self-consistent calculations
demonstrate a breakdown of the SCBA for large $\rm{\alpha}$ and we identify a weak and strong coupling regime. 
For weak coupling our SCBA results compare closely with those obtained with exact scattering
theory. However in the strong coupling regime large deviations are found. In particular we demonstrate that there is a critical
coupling strength, characteristic of the  materials system, beyond which multiple self-consistent solutions 
can be found depending on the initial conditions in the simulation. We attribute these features to the breakdown of
the perturbative expansion leading to the SCBA.

\end{abstract}

\maketitle

%*********************************************************************
% Introduction
%*********************************************************************
\section{\label{sec:intro}Introduction}

Central to the field of molecular electronics are phenomena involving the interaction between the electron current 
and the internal degrees of freedom of the molecule investigated. In scanning tunnel microscopy (STM) 
\cite{Pascual1, ohara,Komeda} the molecular vibrational modes (phonons) have been exploited to desorb 
or to move a molecule on a surface, paving the way for phonon assisted surface chemistry. At the same time,  STM 
inelastic tunnelling spectroscopy uses the fingerprints of vibrations in the $I$-$V$ curve to probe the orientation 
and/or to identity molecules on surfaces \cite{Sainoo,Hipps, Stipe2,Lambe}. Switching devices exploiting phonons have 
also been reported \cite{Mayer}.

Broadly speaking, in molecular devices phonons are important for two reasons. Firstly, they
play a role in transport\cite{Smit,Agrait} by opening new conductance channels through which the itinerant electrons can
propagate, and by suppressing the transmission of purely elastic channels \cite{Nicola}. More dramatically, for large 
electron-phonon coupling the charge carriers become quasi-particles consisting of coupled electrons and phonons \cite{Ness}. 
Secondly, from a technological point of view, phonons limit the efficiency of molecular devices because of energy dissipation. 
This causes heating, power loss and instability.

Transport experiments at the nano-scale are difficult to interpret since the atomically precise device geometry is
rarely known. Therefore one usually relies on atomistic simulation techniques in order to understand the results. 
For elastic transport, when electron-phonon interaction is not considered and the electron-electron interaction is treated 
at the mean field level, methods of note for predicting the current flowing through devices include the 
non-equilibrium Green function formalism (NEGF) \cite{Keldysh,Alex2,Rammer,Daniele,Caroli} and scattering
theory (ST) \cite{Magoga,ButtikerImry,Ness,Lambert}. Some of these methods have been adapted to include
electron-phonon interaction, notably an extension of scattering theory (EST)\cite{Bonca,Bonca2,Emberly,Nicola} 
and the self consistent Born approximation (SCBA)\cite{Freddy1,Galperin1} within the NEGF formalism. In addition 
time-dependent methods for describing correlated electron-ion dynamics have been recently proposed\cite{Todorov}.

The focus of this paper is the SCBA. This is attractive from a practical point of view since it has moderate computational 
requirements and it has been used extensively for calculating transport properties of a number of different material 
systems \cite{Freddy1,Galperin1}. However, it is a perturbative approach appropriate only for weak electron-phonon (e-p) coupling. 
As the e-p coupling strength increases the SCBA will eventually breakdown, however it is unclear whether such breakdown 
is either sharp or smooth with the e-p coupling strength. Our work explores this question in detail.

The paper is organized as follows. We begin by presenting the NEGF formalism for a two probe device\cite{Alex,Datta}, 
and by recalling the foundations of the SCBA. We then consider a 1D tight-binding model where the e-p interaction in the 
scattering region is described by the Su-Schrieffer-Heeger (SSH)\cite{Su1,Su2} Hamiltonian. 
The parameters for the model Hamiltonian are chosen for mimicking two systems which have 
been studied experimentally: $\rm{H_2}$ molcules sandwiched between Pt electrodes ($\rm{H_2}$-Pt)\cite{Smit,Djukic} and 
Au monatomic chains\cite{Agrait} comprising R atoms (RC's). The parameters for $\rm{H_2}$-Pt are the same as those used 
by Jean and Sanvito \cite{Nicola}, who previously employed exact scattering theory (EST) to describe phononic effects. In contrast 
to the SCBA, EST is valid for both strong and weak coupling and therefore it is a good benchmark for the SCBA. Accordingly we 
compare the SCBA results directly with EST over a range of different e-p couplings to establish the limit of validity for the 
SCBA and to investigate its breakdown. 

\section{Methodology}\label{sec:Meth}
\subsection{Non Equilibrium Green Function Formalism }\label{subsec:NEGF}

A two probe device consists of two crystalline electrodes attached on either side of a scattering region, which is in general 
a collection of atoms breaking the electrodes translational symmetry. The leads are also charge reservoirs, so 
that the device may be viewed as two charge reservoirs bridged by the central region. Thermodynamically we characterise 
the left-hand side (L) and right-hand side (R) lead by defining their chemical potentials $\rm{\mu_{L} and~ \mu_{R}}$. 
If $\rm{\mu_{L}=\mu_{R}}$, equilibrium is established and no current flows. When $\rm{\mu_{L}\neq \mu_{R}}$ the system 
is dragged out of equilibrium, and net charge will move from the reservoir with the higher chemical potential across the central 
region to the reservoir of lower chemical potential in an attempt to re-establish equilibrium. If a battery is attached to the two 
reservoirs keeping $\rm{\mu_{L}- \mu_{R}}=eV$ ($V$ is the bias and $e$ the electron charge) the system cannot return to 
equilibrium and will eventually reach a steady state with a constant current flow.

\begin{figure}[hbt]
\begin{center}
\includegraphics[clip=true,width=.48\textwidth]{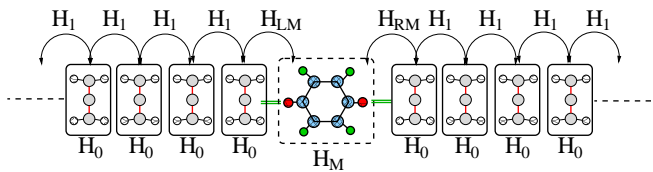}
\end{center}
\caption{\small{Schematic representation of a system composed of two semi infinite leads and a scattering region 
(rectangular dashed box). The matrices $H_0$ and $H_1$ describe the lead principal layers, $H_\mathrm{{M}}$ 
describes the scattering region and $H_\mathrm{LM}$, $H_\mathrm{RM}$ the interaction between the scattering region
and the last principal layers of the leads.}} 
\label{GeneralSystem}
\end{figure}

At the Hamiltonian level the problem can be formulated by using a basis set comprising a linear combination of atomic orbitals 
(LCAO). It is convenient to write the Hamiltonian of the semi-infinite periodic leads in term of principal layers (PLs) \cite{Alex,Alex3}.
These are cells that repeat periodically and constructed in such a way that the interaction between PLs extend only 
to nearest neighbours (see figure \ref{GeneralSystem}).
Thus the $N \times N$ matrices $H_1$ and $H_0$ describe respectively the interactions between PLs
and within a PL. The scattering region in general is described by $M$ basis functions. The $M\times M$ matrix, 
$H_\mathrm{M}$, describes its internal interaction, while the matrices $H_\mathrm{LM}$ ($N\times M$) and $H_\mathrm{RM}$
($M\times N$) contains the interaction between the PLs of the leads adjacent to the scattering region and the scattering 
region itself. The entire system is thus described by the infinite tri-diagonal Hamiltonian $\mathrm{\cal{H}}$ 
\begin{equation}
 \cal{H} = \mathrm{\left( \begin{array}{ccccccccccc}
           .&.&.&.&.&.&.&.&.&.&. \\
	  .&0&H_{-1}&H_0&H_1&0&.&.&.&.&. \\
	.&.&0&H_{-1}&H_0&H_\mathrm{LM}&0&.&.&.&. \\
	.&.&.&0&H_\mathrm{ML}&H_\mathrm{M}&H_\mathrm{MR}&0&.&.&. \\
	.&.&.&.&0&H_\mathrm{RM}&H_0&H_1&0&.&. \\
	.&.&.&.&.&0&H_{-1}&H_0&H_1&0&. \\
	.&.&.&.&.&.&.&.&.&.&. \\
	\end{array}\right)}\;.\nonumber
\end{equation} 
Time reversal symmetry sets $H_{-1}= H_{1}^{\dag}$,  $H_\mathrm{ML}= H_\mathrm{LM}^{\dag}$, and 
$H_\mathrm{MR}=H_\mathrm{RM}^{\dag}$. The retarded Green function, ${\cal{G}}^\mathrm{R}$
associated to the entire system (leads plus scattering region) is defined as
\begin{equation}
 \mathrm{[\omega^{\prime}\cal{I} - \cal{H}]{\cal{G}}^\mathrm{R}(\omega) = \cal{I},}
\end{equation}
where $\rm{\omega^{\prime} = lim_{\delta \rightarrow 0^{+}}~ (\omega + i \delta})$, $\rm{\omega}$ is the energy and 
$\cal{I}$ is the infinite dimensional identity matrix.\footnote{The superscript ``R'' is omitted throughout the paper
unless necessary.}

For transport calculations however one does not need the Green's function of the entire system but only that
relative to the scattering region, $G_\mathrm{M}$, in presence of the leads. This can be written as
\footnote{A generalisation to a non-orthogonal LCAO Hamiltonian is described in reference \cite{Alex3}.}
\begin{equation}
G_\mathrm{M}(\omega) = [\omega^{\prime}I_\mathrm{M} - H_\mathrm{M} - \Sigma_\mathrm{L}(\omega) - 
\Sigma_\mathrm{R}(\omega) ]^{-1}\label{GM}\;,
\end{equation}
where the presence of the leads have been accounted via the introduction of the self-energies for the left- and right-hand side 
lead $\rm{\Sigma_{L}(\omega)}$ and $\rm{\Sigma_{R}(\omega)}$. $I_\mathrm{M}$ is the $M \times M$ identity matrix.
The self energies are $M\times M$ matrices defined as
\begin{equation}
\Sigma_\mathrm{L}=H_\mathrm{ML}\:g_\mathrm{L}\:H_\mathrm{LM}\:,
\;\;\;\Sigma_\mathrm{R} = H_\mathrm{MR}\:g_R\:H_\mathrm{RM}
\:,\label{Lself}
\end{equation}
%
%\begin{equation}
%\Sigma_\mathrm{R}(\omega) = H_\mathrm{MR}\:g_R(\omega)\:H_\mathrm{RM}\:,\label{Rself}
%\end{equation} 
%
where $g_\mathrm{L}(\omega)$ and $g_\mathrm{R}(\omega)$ are the retarded \emph{surface} Green functions of the leads, namely 
the retarded Green functions of the isolated semi-infinite leads evaluated at the PLs adjacent to the scattering region. 
These are calculated by considering the retarded Green function of the corresponding infinite system (periodic) and by applying 
appropriate boundary conditions\cite{Lambert}.
External bias voltage is introduced under the assumption that the leads are good metals maintaining local charge
neutrality. The effect of a bias is therefore only that of shifting rigidly in energy the leads electronic structure, so that
\begin{equation}
\Sigma_\mathrm{L/R}(\omega,V)=\Sigma_\mathrm{L/R}(\omega \pm eV/2,0)\:.\label{S1}
%\Sigma_\mathrm{R}(\omega,V)=\Sigma_\mathrm{R}(\omega - eV/2,0)\:.\label{S2}
\end{equation}

We now proceed to evaluate the non-equilibrium charge density in the scattering region and the two-probe current by using the 
NEGF scheme\cite{Alex3}. The lesser ($<$) and greater ($>$) Green functions $G_\mathrm{M}^{\lessgtr}(\omega)$ are defined as 
\begin{eqnarray}
G^{\lessgtr}_\mathrm{M}(\omega)  &=& G_\mathrm{M}(\omega)\Sigma^{\lessgtr}(\omega)G^{\dag}_\mathrm{M}(\omega)\label{Gless},
\end{eqnarray}
with self-energies
\begin{eqnarray}
\mathrm{\Sigma^{\lessgtr}(\omega)} & = & \mathrm{\sum_{\alpha = L, R}\Sigma_{\alpha}^{\lessgtr}(\omega)}\:,\label{selflesst}\\
\Sigma_{\alpha}^{<}(\omega) & = & i\:n_\mathrm{F}^\alpha(\omega)\Gamma_{\alpha}(\omega)\:,\label{selflessl}\\
\Sigma_{\alpha}^{>}(\omega) & = & i\:[n_\mathrm{F}^\alpha(\omega)- 1]\Gamma_{\alpha}(\omega)\:.\label{selfgreatl}
\end{eqnarray}
Here $n_\mathrm{F}^\alpha(\omega)=n_\mathrm{F}(\omega-\mu_\alpha)$ is the Fermi function evaluated at 
${\omega}-\mu_\alpha$ and temperature $T$, $\mu_{\alpha} = E_\mathrm{F} \pm eV/2$ with $E_\mathrm{F}$ 
the leads Fermi energy, and we have introduced the coupling matrix for the ${\alpha}$ lead ($\rm{\alpha=L/R}$) 
\begin{equation}
\Gamma_{\alpha}(\omega,V) = i\rm{[\Sigma_{\alpha}(\omega,V) - \Sigma^{\dag}_{\alpha}(\omega,V)]}\:.\label{Coupling}
\end{equation}
The non-equilibrium charge density matrix for the scattering region is
\begin{equation}
\rho= \frac{1}{2\pi i} \int_{- \infty}^{\infty} \mathrm{d}\omega^\prime G_\mathrm{M}^<{(\omega^\prime)}\:.\label{rho}
\end{equation}
If $H_\mathrm{M}$ has a functional dependence on $\rm{\rho}$ the equations (\ref{GM}) and (\ref{rho}) can be solved 
self-consistently. The net current flowing through the device is then
\begin{equation}
J_\mathrm{unp}(V) = \frac{2e}{h}\int_{- \infty}^{\infty}\mathrm{d}\omega~\mathrm{Tr} [\Gamma_\mathrm{L}G^{\dag}_\mathrm{M}
\Gamma_\mathrm{R}G_\mathrm{M}] (n_\mathrm{F}^\mathrm{L}-n_\mathrm{F}^\mathrm{R})\:,\label{elastcurr}
\end{equation}
where the subscript ``{unp}`` stands for ``unperturbed'', meaning that no e-p interaction is included. The term 
$T(\omega,V)=\rm{Tr [\Gamma_{L}G^{\dag}_{M}\Gamma_{R}G_{M}]}$ is the standard Landauer B\"{u}ttiker transmission 
coefficient, although in this case it is explicitly bias dependent. The conductance $G$ in the linear response limit is
\begin{equation}
G = \frac{2e^2}{h}T(E_ \mathrm{F},0),\label{0G}
\end{equation}
while more generally at a given bias $V$ one has
\begin{equation}
G(V) = \left.\frac{\mathrm{d}J_\mathrm{unp}}{\mathrm{d}V}\right|_{V}.\label{Gv}
\end{equation}

\subsection{Self Consistent Born Approximation}\label{subsec:SCBA}

\begin{figure}[hbt]
\begin{center}
\includegraphics[clip=true,width=.45\textwidth]{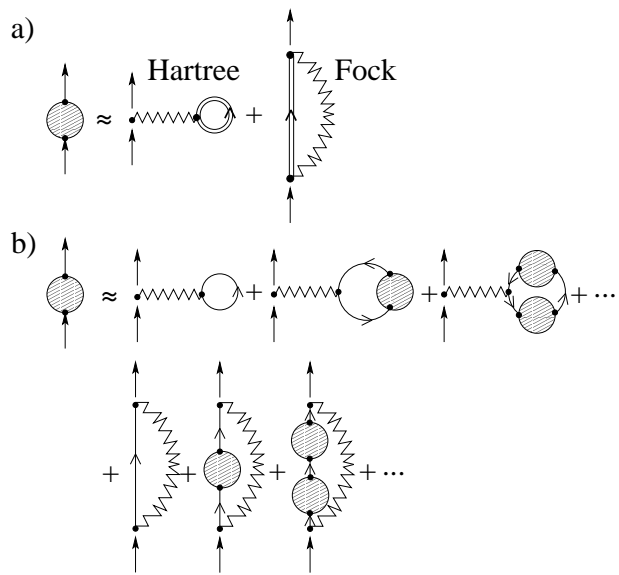}
\end{center}
\caption{\small{Diagrammatic representation of the Hartree-Fock approximation. (a) the self-consistent proper 
self-energy (the shaded circle) is obtained from the first-order Hartree-Fock diagrams evaluated using the interacting 
Green's function (double line). This is equivalent to re-summing all the diagrams\cite{Walecka} in (b).}}
\label{Feynman}
\end{figure}

We now discuss the main concepts associated with introducing e-p scattering into the NEGF transport scheme. 
In general, inelastic scattering produces loss of phase coherence, similarly to what happens when an electron
is absorbed by a reservoir. In fact one may think of inelastic processes as resulting from the coupling of
the scattering region to a ``fictitious'' charge reservoir \cite{Buttiker} that does not exchange a net current.
Thus e-p interaction can be introduced via a self-energy $\rm{\Sigma_{ph}(\omega)}$ and the retarded Green function
becomes
\begin{equation}
G_\mathrm{M}(\omega)  = [\omega^{\prime}I_\mathrm{M} - H_\mathrm{M} - \Sigma_\mathrm{L}(\omega)-
\Sigma_\mathrm{R}(\omega)-\Sigma_\mathrm{ph}(\omega) ]^{-1}\:.\label{Grp}
\end{equation}
The exact form for $\rm{\Sigma_{ph}(\omega)}$ is unknown, however convenient approximations can be derived
from the perturbative expansion over the e-p coupling strength \cite{Daniele,Freddy1,Galperin1,Datta,Jauho}.
In this work we consider the SCBA where only the Hartree and Fock diagrams of the perturbative expansion are retained 
(see figure \ref{Feynman}). This is equivalent to evaluating the first order diagrams at the interacting electronic Green 
function. Thus the phonon self-energy reads 
\begin{equation}
\mathrm{\Sigma_{ph}(\omega) = \Sigma^{F}(\omega) + \Sigma^{H}},
\end{equation}
where the retarded Hartree (H) and Fock (F) contributions to the self-energies are respectively \cite{Datta,Freddy2}
\begin{eqnarray}
\Sigma^\mathrm{H} & = & i\sum_\lambda\frac{4}{\Omega_\lambda} \int_{-\infty}^{\infty} \frac{\mathrm{d}\omega^{\prime}}{2\pi} 
M^{\lambda} \mathrm{Tr}[G_{M}^<(\omega^{\prime})M^{\lambda}]\:,\label{Hart} \\
\mathrm{\Sigma^{F}(\omega)}  & = & \mathrm{\sum_{\lambda}\Biggl[\frac{1}{2}[\Sigma_{\lambda}^{>}(\omega) - \Sigma_{\lambda}^{<}(\omega)]}\nonumber\\
&& -\mathrm{\frac{i}{2}{\cal H}_{\omega^{\prime}}\{\Sigma_{\lambda}^{>}(\omega^{\prime}) - \Sigma_{\lambda}^{<} (\omega^{\prime})\}(\omega)\Biggr]}.\label{Fock}
\end{eqnarray}
In equation (\ref{Fock}) ${\cal{H}}_{\omega^{\prime}}$ is the Hilbert transform
\begin{equation}
{\cal H}_{x}\{f(x)\}(y) = \frac{1}{\pi}{\cal P}\int_{- \infty}^{\infty} dx\frac{f(x)}{x - y},
\end{equation}
and ${\cal P}$ stands for the principal part of the integral. The phonon energy and e-p coupling matrix for a particular mode $\rm{\lambda}$
are respectively $\rm{\Omega_{\lambda}}$ and $M^\mathrm{\lambda}$. Finally the e-p lesser and greater self-energy 
are given by
\begin{eqnarray}
\mathrm{\Sigma^{F\lessgtr}(\omega) } & = & \mathrm{\sum_{\lambda}\Sigma_{\lambda}^{\lessgtr}(\omega)},\label{GrtrLssSCBA}\\
\Sigma_{\lambda}^{\lessgtr}(\omega)  & = &M^{\lambda} \Biggl[\left(N_{\lambda}+1 \right)G_\mathrm{M}^\lessgtr(\omega \pm\Omega_{\lambda}) \nonumber\\&+& N_{\lambda}G_\mathrm{M}^{\lessgtr}(\omega\mp\Omega_{\lambda})\Biggr]M^{\lambda}\:,\label{LesserPh}
\end{eqnarray}
which is simply a sum of the lesser self-energies over all the possible modes $\rm{\lambda}$. The occupancy of each phonon 
mode is ${N_{\lambda}}$. 

We assume that the phonons are in thermal equilbrium with a bath so that for a given temperature ${N_{\lambda}}$ is simply the 
Bose-Einstein distribution $N_{\lambda} = (e^{\Omega_\lambda/K_\mathrm{B}T} - 1)^{-1}$, with ${K_\mathrm{B}}$ the Boltzman 
constant. From equation (\ref{LesserPh}) we note that the lesser (greater) self-energy contains contributions from two scattering 
processes: electrons with energy $\rm{\omega - \Omega_{\lambda}}$ may absorb (emit) a phonon and/or electrons with 
energy $\rm{\omega + \Omega_{\lambda}}$ may emit (absorb) a phonon of energy $\rm{\Omega_{\lambda}}$. When
$T\sim 0$, electrons may only emit phonons since $N_\lambda\sim 0$, i.e. no phonons are present in the scattering 
region provided that the phonon lifetime is much smaller than that of the electrons. The total lesser self-energy of equation (\ref{selflesst}) 
must be adjusted to include the phonon lesser self-energy,
%
%\begin{equation}
%\mathrm{\Sigma^{\lessgtr}(\omega) = \Biggl(~\sum_{\alpha =~ L,R} \Sigma^{\alpha\lessgtr}(\omega)\Biggr) + \Sigma_{ph}^{\lessgtr}(\omega)},\label{selflesspt}
%\end{equation}
%
%
\begin{equation}
\mathrm{\Sigma^{\lessgtr}(\omega) = \sum_{\alpha =~ L,R} \Sigma^{\alpha\lessgtr}(\omega) + \Sigma_{ph}^{\lessgtr}(\omega)}\label{selflesspt}
\end{equation}
and $\rm{G^{\lessgtr}(\omega)}$ from equation (\ref{Gless}) are now evaluated using the perturbed Green function and lesser 
self-energy from equations (\ref{Grp}) and (\ref{selflesspt}). The general expression for the interacting (including e-p coupling) 
current\cite{Meir} through lead $\rm{\beta}$ may be written as the sum of an elastic and an inelastic contribution 
\begin{eqnarray*}
{J^{\beta}(V)} & = &{J^{\beta}_ \mathrm{el}(V)} + {J^{\beta}_ \mathrm{inel}(V)}\label{PCurr},
\end{eqnarray*}
where
\begin{equation}
J^{\beta}_ \mathrm{el} ={\frac{2e}{h}\int_{- \infty}^{\infty} \mathrm{d}\omega\:  \mathrm{Tr} [\Gamma_{\beta}
G_ \mathrm{M}\Gamma_{\alpha}G_ \mathrm{M}^{\dag}]}
{(n_ \mathrm{F}^{\beta} - n_\mathrm{F}^{\alpha})}
\label{Elasp}
\end{equation}
and
\begin{eqnarray}
J^{\beta}_ \mathrm{inel} = {\frac{2e}{h}\int_{- \infty}^{\infty} \mathrm{d}\omega\:  \mathrm{Tr} \Biggl[\Sigma^{<}_\beta
G_ \mathrm{M}\Sigma_\mathrm{ph}^{>}G_ \mathrm{M}^{\dag}}
- {\Sigma^{>}_\beta G_ \mathrm{M}\Sigma_\mathrm{ph}^{<}G_ \mathrm{M}^{\dag}\Biggr]}.\label{Inelasp}
\label{InelasticJ}
\end{eqnarray}

\section{Numerical Method}\label{sec:Num}
\subsection{Model Hamiltonian and Coupling Matrices ${M^{\lambda}}$}\label{subsec:SSHHam}

The systems under investigation are 1D linear atomic chains described by a $s$-orbital nearest-neighbour
tight-binding model. The scattering region comprises $R$ atoms plus one PL (one atom) from 
each lead, so that it contains ${M=R+2}$ orbitals. Henceforth we refer to this system as RC. 
Furthermore, we assume that the two leads are identical. 
\begin{figure}[hbt]
\begin{center}
\includegraphics[clip=true,width=.48\textwidth]{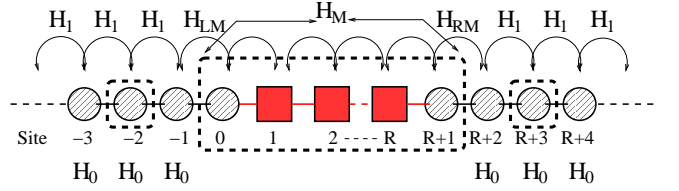}
\end{center}
\caption{\small {Schematic diagram of the simple monatomic systems considered here. It is composed of two semi infinite 
leads and a scattering region. The scattering region is marked with a dashed rectangle. Inelastic scattering is effective only 
in the scattering region.}}
\label{Simplesystem}
\end{figure}
The matrices ${H_{0},~ H_{1}}$ for a 1D tight-binding model reduce to c-numbers: $\epsilon_\mathrm{L}=H_{0}$ and
$\gamma_\mathrm{L}=H_{1}$, where $\rm{\epsilon_L,\gamma_L}$ are the lead onsite energy and hopping parameter 
respectively. The leads' Hamiltonians thus read
\begin{eqnarray}
{{\cal{H}}_\mathrm{L}}&=&{\epsilon_\mathrm{L}\sum_{i = -\infty}^{-1}c^{\dag}_{i}c_{i} +\gamma_\mathrm{L}\sum_{i = -\infty}^{-2}[c^{\dag}_{i}c_{i+1} + c^{\dag}_{i+1}c_{i}]\:,}\\
{{\cal{H}}_{R}}& = &{\epsilon_\mathrm{L}\sum_{i = R+2}^{\infty}c^{\dag}_{i}c_{i} +\gamma_\mathrm{L}\sum_{i = R+2}^{\infty}[c^{\dag}_{i}c_{i+1} + c^{\dag}_{i+1}c_{i}]\:,}
\end{eqnarray}
where $\{{ c^{\dag}_{i},c_{i}}\}$ are the electronic creation and annihilation operators at site $i$. The interaction Hamiltonian 
between the leads and the scattering region is 
\begin{eqnarray}
\mathrm{{\cal{H}}_{LM}}&=&{\gamma_\mathrm{LM}[c^{\dag}_{-1}c_{0} + c^{\dag}_{0}c_{-1}]}\\
\mathrm{{\cal{H}}_{RM}}& = &{\gamma_\mathrm{LM}[c^{\dag}_{R+1}c_{R+2} + c^{\dag}_{R+2}c_{R+1}]}\:,
\end{eqnarray}
where for our setup $\rm{\gamma_{LM}=\gamma_L}$.

For a 1D system the lead self-energies are analytical
\begin{equation}
{\Sigma^\mathrm{R}_{L(R)}(\omega) = \Delta(\omega) - i\frac{\Gamma(\omega)}{2}},
\end{equation}
where
\begin{eqnarray}
\Delta(	\omega) & =  &\rm{\Gamma_{0}}\left\{\begin{array}{ll}
 {x},&  {|x| \leq 1}\\
 {x -\sqrt{x^2 -1}}& {|x| > 1}
\end{array}\:,
\right.\\
\rm{\Gamma(\omega)}& = & {-2\Gamma_{0}\theta(1 - |x|)(\sqrt{1- x^2})}\\
\mbox{and}& &\Gamma_{0} = {\frac{{\gamma_\mathrm{LM}}^2}{\gamma_\mathrm{L}},~~~~~~~x =
 \frac{(\omega - \epsilon_\mathrm{L})}{2\gamma_\mathrm{L}}}.\nonumber
\end{eqnarray}
Finally e-p interaction is included into the scattering region in the form of an SSH Hamiltonian \cite{Su1,Su2}, comprising three terms
\begin{equation}
H_\mathrm{M} = H_\mathrm{e} + H_\mathrm{ep} + H_\mathrm{ph}\:,\label{SSfinal}
\end{equation}
with
\begin{eqnarray}
H_\mathrm{e} & = & {\sum_{i=0}^{R+1}\epsilon_i c_{i}^{\dag} c_{i} + \sum_{i \neq j }\gamma_{ij}^0(c_{i}^{\dag} 
c_{j}+c_{j}^{\dag} c_{i}),\label{elH}}\\
{H_\mathrm{ph}}&=&{\sum_{\lambda=1}^{\lambda_\mathrm{max}} (b_{\lambda}^{\dag} b_{\lambda} + \frac{1}{2})\hbar\omega_{\lambda}} ,\label{pH}\\
{H_\mathrm{ep}} & = &{\sum_{\lambda=1}^{\lambda_\mathrm{max}}\sum_{i \neq j}M_{ij}^{\lambda}(b_{\lambda}^{\dag}+b_{\lambda})(c_{i}^{\dag} c_{j}+c_{j}^{\dag}
c_{i}).\label{epH} }
\end{eqnarray}
$H_\mathrm{e}$ is the electronic Hamiltonian of the scattering region with onsite energies ${\epsilon_i}$ and the unperturbed hopping parameters 
${\gamma^0_{ij}}$ (we assume ${\gamma^0_{ij}=0}$ for ${j \ne i \pm 1}$). ${H_\mathrm{ph}}$ is the non-interacting phonons Hamiltonian, 
written in terms of the phononic creation and annihilation operators ${\{b^{\dag}_{\lambda},b_{\lambda}\}}$ and the phonon 
energies ${\Omega_{\lambda}=\hbar \omega_{\lambda}}$, with the index $\rm{\lambda}$ running over all the modes
\footnote{The two lowest energy modes correspond to the trivial translation of the whole system, and to a fixed chain 
with slowly oscillating electrodes. These are of no interest and have been discarded. The remaining phonon energies have been 
labelled in order of increasing energy such that $\rm{\lambda=1}$ corresponds to the lowest energy mode.}. 
The final term, ${H_\mathrm{ep}}$, is the Hamiltonian describing the e-p interaction within the scattering region. The details of such
interaction are included in the e-p coupling matrices ${M^{\lambda}_{ij}}$.

In order to calculate the matrices ${M^{\lambda}_{ij}}$ and the longitudinal phonon frequencies we consider a simple
nearest neighbours elastic model \cite{Emberly,Freddy1,Kittel} in which
\begin{eqnarray}
{M_{ij}^{\lambda}=\alpha_{ij}\left\{\frac{e_{\lambda}^i}{\sqrt{m_i}} -
\frac{e_{\lambda}^j}{\sqrt{m_j}}\right\} \sqrt{\frac{\hbar}{2 \omega_{\lambda}}}}.
\label{epcouplmat}
\end{eqnarray}
In equation (\ref{epcouplmat}) the orthonormal vectors ${e}_{\lambda}$ represent the ionic displacement associated to 
each mode $\rm{{\lambda}}$, ${m_i}$ is the mass of the atom at site ${i}$, and the constants ${\alpha_{ij}}$ are the e-p 
coupling parameters. These latter are defined as the first order coefficient of the expansion
of the tight-binding hopping parameter $\gamma_{ij}$ about the atomic equilibrium positions
\begin{equation}
{\gamma_{ij}=\gamma^{0}_{ij}+\alpha_{ij}(u_i-u_j)\:,}
\label{SSHlexp}
\end{equation}
where $u_i$ is the displacement vector of the atom at site $i$. From equation (\ref{SSHlexp}) it follows that ${\alpha_{ij}=-\alpha_{ji}}$. 
In the nearest neighbour approximation the interaction is restricted to electrons moving between sites $i$ and ${i \pm 1}$. Finally, 
we note that the eigenvectors ${e_{\lambda}}$ are real. This implies that the matrices ${M^{\lambda}_{ij}}$ are real and symmetric 
with non-zero matrix elements for ${i = i \pm 1}$, so that ${M^{\lambda}_{ij}}$ has a tri-diagonal form for longitudinal phonons. We 
note that, although it is possible calculate the coupling parameters ${\alpha_{ij}}$ using first-principles electronic structure methods, 
here we set ${\alpha_{ij}=\alpha}$ and $\rm{\alpha}$ is taken as a free parameter.

\subsection{Numerical Integration and Self Consistency}\label{subsec:sc}

\begin{figure}[hbt]
\begin{center}
\includegraphics[clip=true,width=.41\textwidth]{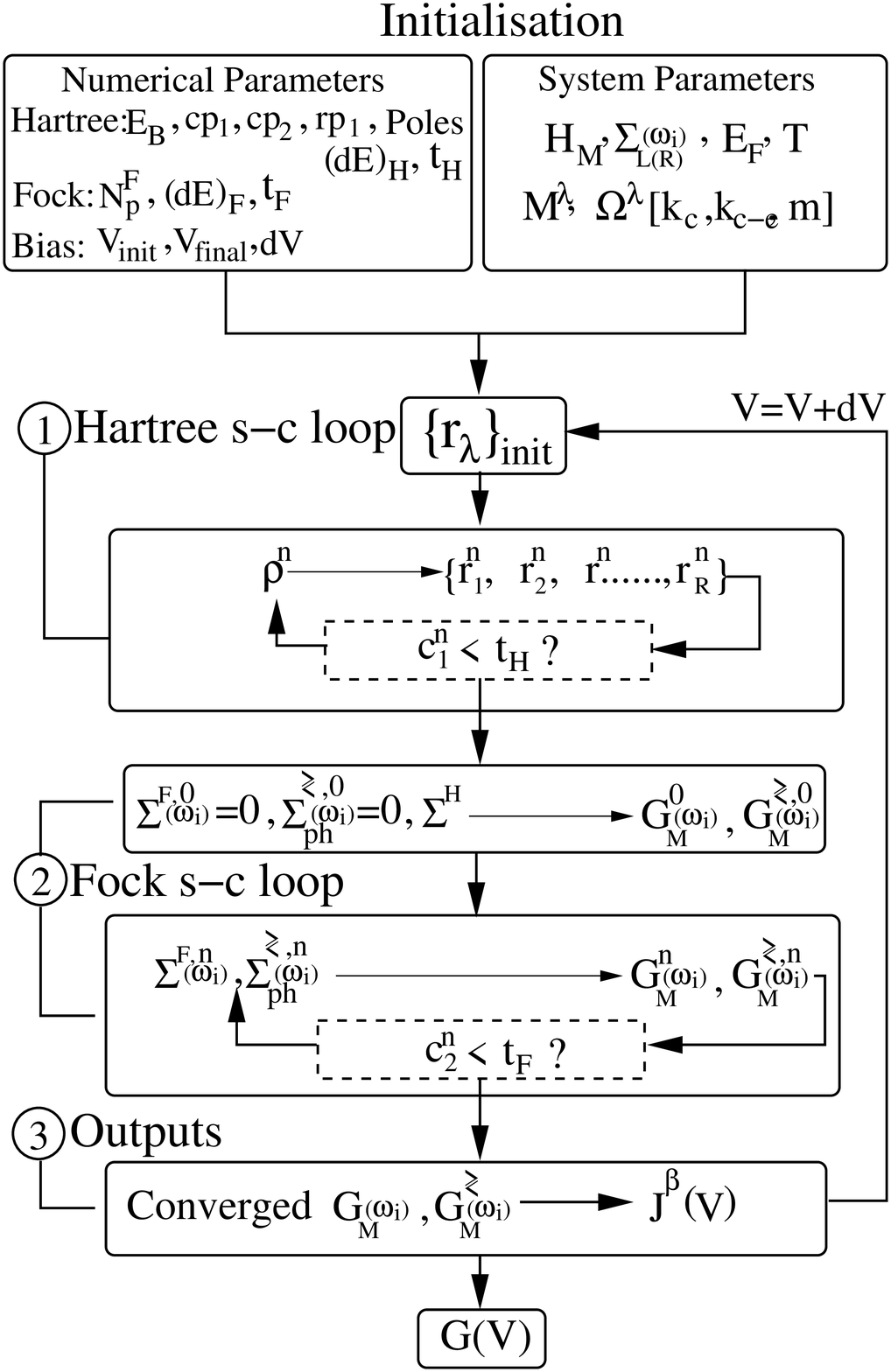}
\end{center}
% \begin{center}
\caption{\small{Scheme of our numerical procedure for self-consistency.}}
\label{selfconp}
\end{figure}
The flowchart in figure \ref{selfconp} outlines the numerical procedure used to calculate the interacting current ${J^{\beta}(V)}$ 
and the differential conductance $G(V)$. Each simulation can be partitioned into three steps. The first two consist of two 
self-consistent loops which calculate the phonon self-energies $\Sigma^\mathrm{H}$ and $\Sigma^\mathrm{F}$ respectively. 
These are used in the third step to evaluate ${J^{\beta}(V)}$ by using the equations (\ref{Elasp}) and (\ref{Inelasp}). 

Let us now discuss the three steps in some detail. We start by writing $\Sigma^\mathrm{H}$ as an explicit function of the 
density matrix $\rm{\rho}$
\begin{eqnarray}
\Sigma^\mathrm{H}(\rho) 
&=&{-4}\sum_{\lambda}\mathrm{Tr}\biggl[{\rho}\frac{{M}^\lambda}{\Omega_\lambda}\biggr] {M}^{\lambda}\:, \label{Hdensity}
\end{eqnarray}
where we assume that all the elements of $G^<_\mathrm{M}$ are integrable. $\Sigma^\mathrm{H}$ is thus nothing but a 
weighted sum of the matrices ${M^{\lambda}}$. This can be written in the form
\begin{equation}
\Sigma^\mathrm{H}=\sum_{\lambda=1}^M r_{\lambda}R^{\lambda}\:,\label{RH}\\
\end{equation}
where the ratio matrices ${R^{\lambda}}$ and their weighting coefficients ${r_{\lambda}}$ are given by
\begin{eqnarray}
R^{\lambda}&=&\frac{|{\gamma^{0}}|_\mathrm{min}}{|M^{\lambda}|_\mathrm{max}}M^{\lambda}.\label{RatM}\\
r_{\lambda} & = &  -{\frac{4|M^{\lambda}|_\mathrm{max}}{\Omega^{\lambda}|{\gamma^{0}}|_\mathrm{min}}\sum_{i=1}^{M-1} 
R^{\lambda}_{i,i+1}(\rho_{i,i+1}+\rho_{i+1,i})},\label{Hweightsgeneral}
\end{eqnarray}

For a given mode ${\lambda}$, the largest matrix element of the matrix ${M^{\lambda}}$ is denoted as ${|M^{\lambda}|_\mathrm{max}}$.
${|\gamma^{0}|_\mathrm{min}}$ is the smallest among the hopping parameters of the unperturbed system (no e-p coupling),
which, by construction, is equal to $|R^{\lambda}|_\mathrm{max}$. The matrices $R^{\lambda}$ are independent of the e-p coupling 
$\alpha$ and simply reflect the symmetry of the specific phonon mode considered. Thus ${r_{\lambda}}$ measures the maximum fractional 
modification of the elements ${\gamma^0_{ij}}$ in the electronic Hamiltonian as the result of e-p coupling.

The first self-consistent loop of  figure \ref{selfconp} begins by choosing initial values for the weighting coefficients,
$\{r_{\lambda}\}_\mathrm{init} = \{r_1^{0}, r_2^{0}, ..., r_R^{0} \}$, which are used to construct the density matrix at the first 
iteration, ${\rho^1}$. Then both ${\{r_{\lambda}\}}$ and ${\rho}$ are varied until the convergence condition
\begin{eqnarray}
{c^n_1 = \sum_{\lambda}\Biggl(\frac{|r^n_{\lambda} -r^{n-1}_{\lambda}|}{|r^{n-1}_{\lambda}|} \Biggr)}&<&t_\mathrm{H}\label{ConvH}
\end{eqnarray}
is met for the chosen tolerance ${t_\mathrm{H}}$. The density matrix $\rho$ is obtained by integrating $G^{<}$ as in equation
(\ref{rho}). In order to perform this integral we first take $\rm{\Sigma^{F} = 0}$ under the assumption that this has little effect
of the convergence of ${\rho}$ (first self-consistent loop). Following previous works \cite{Alex,Brandbyge} we write 
${\rho = \rho_\mathrm{eq} + \rho_\mathrm{V}}$, where
\begin{eqnarray}
\mathrm{\rho_{eq}} & = & -\frac{1}{\mathrm{\pi}} \int_{-\infty}^{\infty} \mathrm{d\omega}\:\mathrm{Im}[G_\mathrm{M}]\:
n_\mathrm{F}^\mathrm{L}\:,\label{Deq}
\end{eqnarray}
and 
\begin{equation}
\rho_\mathrm{V}=\frac{1}{2\pi} \int_{-\infty}^{\infty} \mathrm{d\omega}\:G_\mathrm{M}
\Gamma_\mathrm{R}G_\mathrm{M}^{\dag}\:(n_\mathrm{F}^\mathrm{R}-n_\mathrm{F}^\mathrm{L})\:.
\label{Dv}
\end{equation}
At equilibrium ($\rm{\mu_L=\mu_R=\mu}$) one has $\rm{\rho = \rho_{eq}}$.
\begin{figure}[hbt]
\begin{center}
\includegraphics[clip=true,width=.48\textwidth]{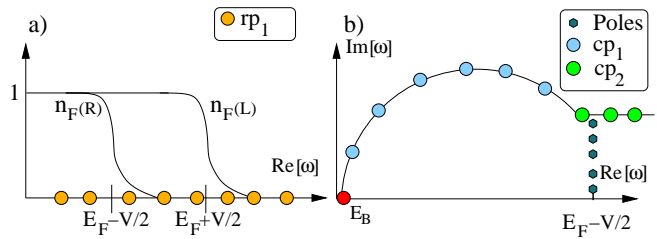}
\end{center}
\caption{\small{Schematic representation of the integrals in equations (\ref{Dv}) and (\ref{Deq}) respectively. In a) the 
integration of $\rm{\rho_V}$ is bound between $E_\mathrm{F} + eV/2$ and $E_\mathrm{F} - eV/2$ by the Fermi 
functions. $\rm{rp_{1}}$ is the number of points of the real axis energy grid. For $\rm{\rho_{eq}}$ in b) the number of 
grid points along the circular path ($\rm{cp_1}$) and the path in the complex plane parallel to the real axis ($\rm{cp_2}$) 
must be chosen, as well as the poles in the Fermi functions.}}
\label{Alex}
\end{figure}

As shown in figure \ref{Alex}a, the integral of $\rm{\rho_V}$ (Eq.~(\ref{Dv})) is along the real axis and it is bound 
between the chemical potentials $\rm{\mu_L}$ and $\rm{\mu_R}$. This is carried out over a numerical grid of sufficient 
fineness $\rm{(dE)_H}$. The calculation of $\rm{\rho_{eq}}$ involves an unbound integral. This is performed over 
a coarse grid in the complex plane using a contour integral method \cite{Williams}, since $G_\mathrm{M}$ is analytical\cite{Brown} 
and smooth in the imaginary energy plane. As shown in  figure \ref{Alex}b a number of numerical parameters must be chosen. 
First the lower limit of integration $E_\mathrm{B}$ must lie below the lowest lying molecular states and below the lowest 
electrode bands. Secondly, the poles of the Fermi functions (Matsubara frequencies) 
which lie within the contour must be taken into account. The integration is then performed by using Gaussian 
quadrature\cite{Vetterling} .

The second self-consistent loop begins by calculating ${\{G^0_\mathrm{M},G^{\gtrless,0}\}}$ using equation (\ref{Grp}), 
where the converged ${\Sigma^\mathrm{H}}$ from the first loop is used and ${\Sigma^\mathrm{F}=0}$. We then proceed to 
iterate ${\{\Sigma^\mathrm{F},\Sigma_\mathrm{ph}^{\gtrless}\}}$ and ${\{G_\mathrm{M},G^{\gtrless}\}}$ until a second 
convergence condition is met
\begin{eqnarray}
{ c^n_2 = \frac{1}{N^\mathrm{F}_\mathrm{p}}\mathrm{Max}\Biggl\{ |[G^{n}_\mathrm{M}(\omega_{i}) -G^{n-1}_\mathrm{M}
(\omega_{i})]|\Biggr\}} &<& t_\mathrm{F}\:.\label{Conv2}
\end{eqnarray}
Note that the condition is over the largest of the matrix elements and runs over the $N^\mathrm{F}_\mathrm{p}$ energy points ${\omega_i}$ 
of the entire grid. Note also that the tolerance used, $t_\mathrm{F}$, is in general different to that used for the Hartree term. 
The Hilbert transform required for calculating the imaginary part of $\rm{\Sigma^F}$ (eq.~(\ref{Fock})) is done by using 
a convolution method combined with a fast Fourier transform algorithm\cite{Freddy1}. In order to avoid end-point corrections we
choose a grid of sufficient range while the grid fineness $\rm{(dE)_F}$ must be sufficiently fine to resolve phononic 
features which lie in the meV range.

Table \ref{SystemParameters} shows the numerical and system parameters used in our simulations. The parameters 
for the $\rm{H_2}$-Pt junctions are identical to those used by Jean\cite{Nicola} within the EST treatment of phonons. This set 
produces the same unperturbed $G(0)\sim .97 G_0$. The spring constants and masses are chosen to give longitudinal 
phonon modes of energies 63~meV (CM mode) and 432~meV respectively while the ratio matrices for these two modes are
\begin{equation}
 {R^1} = \rm{\left( \begin{array}{cccc}
           0&3.2&0&0 \\
	3.2&0&0&0 \\
	0&0&0&-3.2 \\
	0&0&-3.2&0 \\	
	\end{array}\right)}\label{R1}
\end{equation} 
and
\begin{equation}
 {R^2} = \rm{\left( \begin{array}{cccc}
           0&1.6&0&0 \\
	1.6&0&-3.2&0 \\
	0&-3.2&0&1.6 \\
	0&0&-1.6&0 \\	
	\end{array}\right).}\label{R2}
\end{equation}
The e-p coupling $\rm{\alpha}$ remains a free parameter.

The parameters for the Au RC's match closely those used by Frederiksen of \cite{Freddy1}. The atoms in the leads are 
chosen as identical to the atoms in the scattering region, thus that a single onsite energy and hopping parameter 
characterise the electronic Hamiltonian. These parameters and the equilibrium potential $\rm{\mu_{eq}}$ are chosen 
so that the differential conductance for the unperturbed system is ${G_0 = 2e^2/h}$ (perfect transmission).
\begin{table}
\begin{center}
\begin{tabular}{l l l l}
\hline
$ \textrm{ System~parameter} $&$\rm{H_2-Pt} $ &$\rm{Au~ RC} $ &\\
$ \textrm{ Symbol} $ &$\textrm{Value} $ &$\textrm{Value} $ &$ \textrm{Units} $\\
\hline
% $\rm{eV/\AA{~}^2}$ 
$E_\mathrm{F}$ & 0.00 &0.00 & \textrm{eV} \\
$\rm{\epsilon_{M}}$ \textrm{(molecule)}  & -6.0 &  0.0 &\textrm{eV} \\
$\rm{\epsilon_{L}}$ \textrm{(leads)} & 0.00 & -1.00 &\textrm{eV} \\
$\rm{\gamma_{L}}$ \textrm{(leads)} & 5 &  -1.00 &\textrm{eV} \\
$\rm{\gamma_{M}}$ \textrm{(molecule)} & 6.0 &  -1.00 &\textrm{eV} \\
$\rm{\gamma_{LM}}$ & 3.2 & -1.00 &\textrm{eV} \\
T & 4.0 & 4.0 &\textrm{K} \\
$m$ \textrm{(atomic mass)}  & 1  &  197  &\textrm{amu} \\
$\rm{K_{c}}$ & 21.82 & 2.00 &\textrm{eV/\AA}$^2$ \\
$\rm{K_{c-e}}$ & 0.91 & 1.00 &\textrm{eV/\AA}$^2$ \\
\hline 
$ \textrm{ Numerical~parameter} $ &$\textrm{Value} $ &$\textrm{Value} $ &$ \textrm{Units} $\\
\hline
${[V_\mathrm{ini},V_\mathrm{final}]}$ (\textrm{Bias Range}) & [0,200] & [0,31] &\textrm{meV} \\
\textrm{Number of Bias points}  & 250 & 120 &~--~ \\
$\mathrm{(dE)_{F}}$ \textrm{(Fock)}  & $0.1$ & $0.1$ &\textrm{meV} \\
$N^\mathrm{F}_\mathrm{p}$ \textrm{(Fock)}   & 12600 & 12600 &~--~ \\
$\rm{rp_1}$, \textrm{(Hartree)} & 4000 & 4000 &~~--~ \\
$\mathrm{(dE)_{H}}$ \textrm{(Hartree)}  & .1061 & .0215 &\textrm{meV} \\
$\rm{cp_1~ (Hartree)}$ & 400 & 200 &~--~ \\
$\rm{cp_2~ (Hartree)}$ & 400 & 200 &~--~ \\
$\rm{E_B~ (Hartree)}$ & -28.0  & -5.0  &\textrm{eV} \\
$\rm{Poles~ (Hartree)}$  & 80  & 80  &~--~ \\
${t_\mathrm{F}}$ (\textrm{Tolerance})  & $9.10^{-8}$  & $9.10^{-8}$  &~$\rm{eV}^{-1}$~ \\
${t_\mathrm{H}}$ (\textrm{Tolerance})  & $1.10^{-8}$  & $1.10^{-6}$  &~--~ \\
\hline
\end{tabular}
\caption{\small{Parameters used to simulate the $\mathrm{H}_2$-Pt junctions and Au RC's. The parameter $\rm{K_c}$ is to the spring 
constant between the atoms in the chain, while $\rm{K_{c-e}}$ is to the spring constant between the molecule and the electrodes. 
The Fock grid and real Hartree grid are symmetric about $E_\mathrm{F}$.}}\label{SystemParameters}
\end{center}
\end{table}

\section{Results: H$_2$-P\lowercase{t} junctions}\label{sec:H2pt}
\subsection{self-consistent simulations}\label{subsec:scsim}

\begin{figure}[bth]
\begin{center}
\includegraphics[clip=true,width=.48\textwidth]{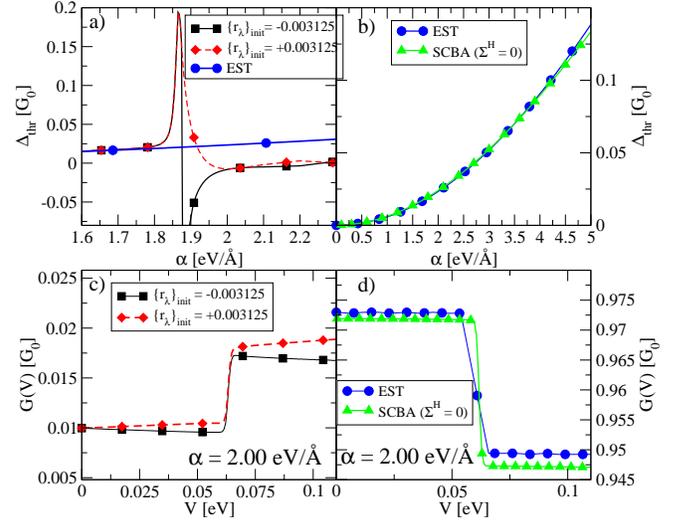}
\end{center}
\caption{\small{Differential conductance and conductance drop at threshold for the H$_2$-Pt junctions - $V_{\rm{thr}}$ is taken at 68.5 meV. Results obtained
using the full SCBA are given for the different initial conditions (panels a and c). In panels b) and d) we also show results 
obtained by setting $\Sigma^\mathrm{H}=0$. In this second case $\Delta_\mathrm{thr}$ and $G(V)$ agree well with the 
EST\cite{Nicola}(EST) up to ${\alpha\approx4.0}$~eV/\AA. The full SCBA disagrees with the EST at a considerably 
lower $\rm{\alpha\approx1.8}$ eV/\AA. Above such a coupling strength $G(V)$ depends on the initial conditions.
}} 
\label{H2FdFull}
\end{figure}

The self-consistent $G(V)$ in the range 0-200~meV for $\alpha$ ranging between 0 and 3.1 eV/\AA\ are presented in 
figure~\ref{H2FdFull}. For this system the characteristic signature of e-p interaction is a drop in the conductance at a 
threshhold voltage $V_\mathrm{thr}$. This signals the onset of inelastic electron processes involving the emission of 
phonons with energies $\Omega\approx eV_\mathrm{thr}$. We quantify this effect by defining the conductance  
drop (in units of $G_0$)
\begin{equation}
 \Delta_\mathrm{thr} = G(0) - G(V_\mathrm{thr})\:.\label{Ddrop}
\end{equation}

Numerical simulations are carried out in two different ways. First, we follow the exact numerical procedure of figure 
\ref{selfconp} (``full SCBA''), but we run simulations starting with different initial conditions, namely 
${\{r_{\lambda}\}_\mathrm{init}=~ + 0.003125}$ and ${\{r_{\lambda}\}_\mathrm{init}=~ - 0.003125}$. 
In figure \ref{H2FdFull}a $\rm{\Delta_{thr}}$ is plotted versus $\rm{\alpha}$ demonstrating good agreement between 
the full SCBA and EST for low $\rm{\alpha}$. However, the two methods disagree for $\alpha$ beyond 
$\rm{\alpha_{crit}\sim 1.8}$ eV/\AA. $\rm{\Delta_{thr}}$ peaks sharply above $\alpha_\mathrm{crit}$,  beyond which 
it becomes dependent on the initial condition ${\{r_{\lambda}\}_\mathrm{init}}$. This last situation is shown 
in figure \ref{H2FdFull}c for $\rm{\alpha=2.0}$ eV/\AA\ where two different $G(V)$ curves are predicted for 
different ${\{r_{\lambda}\}_\mathrm{init}}$ and a low-bias conductance of $0.0125~G_0$ is observed in 
stark disagreement with the unperturbed value of $\sim {0.97~G_0}$.

Since the Hartree self-consistent loop is performed before the Fock one, the dependence on the initial conditions 
suggests that $\Sigma^\mathrm{H}$ may be responsible for the behaviour observed for $\alpha>\alpha_\mathrm{crit}$. 
Stronger evidence to support this hypothesis is provided in figures \ref{H2FdFull}b and \ref{H2FdFull}d where the results 
for our second set of simulations in which $\Sigma^\mathrm{H}$ is set to zero, are presented (``partial SCBA``). 
Figure \ref{H2FdFull}b shows good agreement between the partial SCBA and the EST to the much higher coupling of
$\rm{\alpha \sim 4.0}$ eV/\AA. Moreover there is no evidence of any $\alpha_\mathrm{crit}$. 
This is confirmed by the $G(V)$ curve obtained for $\rm{\alpha=2.0}$ eV/\AA\
and presented in figure \ref{H2FdFull}d.

\subsection{Contribution from the individual modes}\label{subsec:esH}

\begin{figure}[hbt]
\begin{center}
\includegraphics[clip=true,width=.46\textwidth]{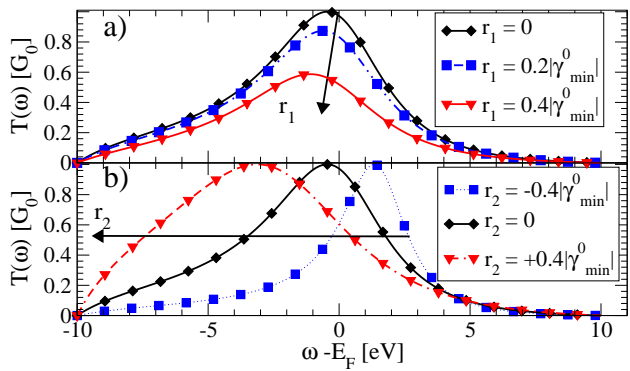}
\end{center}
\caption{\small{Transmission coefficients calculated using the Green function $G_\mathrm{M}^{\lambda}$ of Eq.~(\ref{Gtest}) 
for ${\lambda=1}$ (a) and $\lambda=2$ (b). The matrices ${R^1}$ and ${R^2}$ are fixed and a range of values 
for ${r_{\lambda}}$ has been chosen.}} 
\label{HP2Transguess}
\end{figure}

\begin{figure}[hbt]
\begin{center}
\includegraphics[clip=true,width=.4\textwidth]{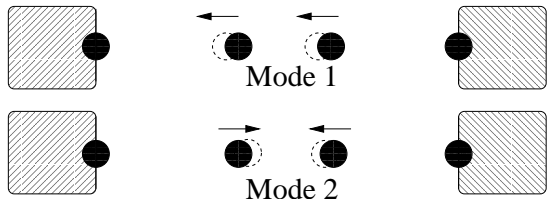}
\end{center}
\caption{\small{The longitudinal vibration modes of H$_2$-Pt junctions. The ${\lambda=1}$ mode of energy 63~meV
is the rigid translational of the H$_2$ centre of mass, while the ${\lambda=2}$ mode is the symmetric mode 
of energy 432~meV. The atoms of the leads are fixed.}}
\label{modediagram}
\end{figure}

We now analyze the origin the breakdown of the full SCBA. For ${V \ll V_\mathrm{thr}}$
the inelastic current $J_\mathrm{inel}$ is strongly suppressed by Pauli exclusion principle. At low temperature 
($T$=4.0~K) we can approximate the Fermi distributions in $J_\mathrm{el}$ by step functions to obtain
\begin{eqnarray*}
\mathrm{J^{\beta}(V)} & \simeq & \frac{2~e}{h}\int_{\mu_\mathrm{L}}^{\mu_\mathrm{R}} T(\omega) \mathrm{d}\omega\:.\label{Japprox}
\end{eqnarray*}
We can now easily probe the contribution of the Hartree term to the conductance by considering the test Green function
\begin{equation}
G_\mathrm{M}^{\lambda}(\omega)=[\omega^{\prime}I_\mathrm{M}-H_\mathrm{M}-\Sigma_\mathrm{L}-\Sigma_\mathrm{R}-
r_{\lambda}R^{\lambda}]^{-1}\:.
\label{Gtest}
\end{equation}
This is used to evaluate the transmission coefficient at $V= 0$ and it is useful to understand the influence of the individual
modes ${\lambda}$ over the transmission. In figure \ref{HP2Transguess} we show $T(\omega)$ as a function of $r_\lambda$
for the two longitudinal modes available in the H$_2$-Pt system (see figure \ref{modediagram}). In the case
of the rigid translational mode ($\lambda=1$ and figure \ref{HP2Transguess}a) $T(\omega)$ is reduced in the region 
around $E_\mathrm{F}$ as ${|r_1|}$ increases. However the general shape of $T(\omega)$ is little affected. This is
somehow expected from the shape of the matrix ${R^1}$ (eq. (\ref{R1})), which indicates that mode 1 causes simply a
change in the hopping parameters ${\gamma^0_{12}}$ and ${\gamma^0_{34}}$ connecting the molecule to the leads.
Importantly when ${\gamma^0_{12}}$ is increased, ${\gamma^0_{34}}$ is reduced by the same amount and vice-versa.
Moreover there is a symmetry $r_1\rightarrow -r_1$.

The results for the symmetric mode $\lambda=2$ are presented in figure \ref{HP2Transguess}b. This time the peak in 
transmission is shifted in energy either to the left or to the right depending on the sign of ${r_2}$. When shifted to the left, 
the peak is broadened while a shift to the right narrows it. In either case the transmission around $E_\mathrm{F}$ is 
reduced.

\begin{figure}[hbt]
\begin{center}
\includegraphics[clip=true,width=.48\textwidth]{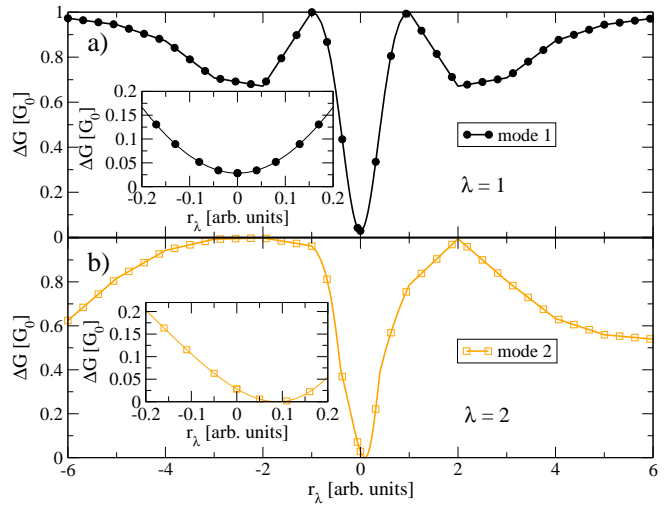}
\end{center}
\caption{\small{Estimate of the deviation ${\Delta G}$ caused by an individual mode ${\lambda}$ as a function of 
${r_{\lambda}}$. The insets shows the region of small perturbation.}}
\label{Hsh}
\end{figure}

From the figures \ref{H2FdFull}c, \ref{HP2Transguess}a, and \ref{HP2Transguess}b one can conclude that the
deviation of the zero bias differential conductance from its unperturbed (no e-p interaction) value is a 
measure of the magnitude of the e-p perturbation. We define this deviation as
\begin{equation}
{\Delta G = G_0 - G(0)}\cong G_0 - T(E_\mathrm{F})\:,\label{DG}
\end{equation}
where the last equality is valid for $T\rightarrow 0$. Figures \ref{Hsh}a and \ref{Hsh}b show the estimated deviation 
(in units of ${G_0}$) versus the weighting coefficients $r_\lambda$. The maximum deviation, $\Delta G =1$, occurs 
when the chain actually breaks as for mode 1 and $r_1 = 1$. As expected, the curve in figure \ref{Hsh}a for mode 1 
is symmetric about 0 while figure \ref{Hsh}b for mode 2 is not.

\subsection{Discussion of the self-consistent results}\label{subsec:revisitH2}

\begin{figure}[hbt]
\begin{center}
\includegraphics[clip=true,width=.48\textwidth]{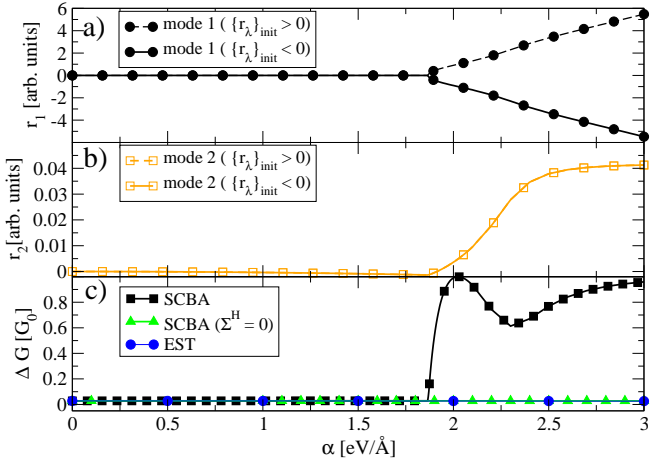}
\end{center}
\caption{\small{Converged self-consistent values of ${r_{\lambda}}$ and deviation $\rm{\Delta G}$ (Eq.~(\ref{DG})) 
as a function of the e-p coupling strength ${\alpha}$ for $V=0$. A transition is apparent at  ${\alpha=1.865 \pm0.005}$~eV/\AA\ 
for the full SCBA. Panels a) and b) show respectively $r_1$ and $r_2$ for the full SCBA for different initial conditions, while in c) 
results for the partial SCBA and EST are also included.}}
\label{rvals}
\end{figure}

We finally re-analyze our self-consistent results in the light of the discussion in the previous section. Figure \ref{rvals}a and \ref{rvals}b 
show the self-consistently calculated ${r_{\lambda}}$ as a function of $\rm{\alpha}$ for $V=0$, while figure \ref{rvals}c shows
$\Delta G$ as defined in equation (\ref{DG}). The critical point ${\alpha_\mathrm{crit}}\sim$1.8~eV/\AA\ marks a sharp transition in the 
behaviour of the weighting coefficients. This is evident in the abrupt change of magnitude and behaviour of ${\Delta G}$ 
(in figure \ref{rvals}c). In fact while for $\alpha<\alpha_\mathrm{crit}$ the full SCBA agrees well with the EST, the two differ
sharply as soon as $\alpha>\alpha_\mathrm{crit}$. Going into more details we note that ${r_1}$ is identically zero
($\sim{10^{-14}}$) for $\alpha<\alpha_\mathrm{crit}$, while $\rm{r_2}$ is small and negative. Importantly both 
${r_1}$ and ${r_2}$ are independent of ${\{r_{\lambda}\}_\mathrm{init}}$. By contrast, for ${\alpha > \alpha_\mathrm{crit}}$, 
${r_2}$ remains independent of ${\{r_{\lambda}\}_\mathrm{init}}$ but this is not the case for $r_1$. In fact we obtain
a strong dependence over the initial conditions with positive (negative) $r_1$ for positive (negative) ${\{r_{\lambda}\}_\mathrm{init}}$.
Importantly none of these features are found in the case when we neglect the Hartree self-energy.

Since ${|r_1|}$ is two orders of magnitude larger than ${|r_2|}$,  it will largely determine $\Sigma^\mathrm{H}$. 
As figure \ref{rvals}a shows, ${r_1}$ varies roughly linearly with ${\alpha}$ above $\alpha_\mathrm{crit}$,
therefore the self-consistent ${\Delta G}$ follows the estimated curve of figure \ref{Hsh}a in this region of ${\alpha}$. 
The magnitude of ${r_1}$ for $\rm{\alpha > \alpha_{crit}}$ suggests that the interaction with phonons becomes a strong 
perturbation of the electronic system. We define the region $\rm{\alpha < \alpha_{crit}}$ as the weak coupling regime and the 
region $\rm{\alpha > \alpha_{crit}}$ as the strong coupling regime.

Figure \ref{rvals} adequately explains the causes of the massive reduction of $G(V)$ with respect to its unperturbed value 
observed in figure \ref{H2FdFull}c at $V=0$. Notably, 
as figure \ref{H2FdFull}a shows, $\Delta_\mathrm{thr}$ calculated with the full SCBA starts deviating from the EST result for 
${\alpha\simeq1.75~}$~eV/\AA, i.e. at a value lower that $\alpha_\mathrm{crit}= 1.865 \pm 0.005$~eV/\AA\ calculated for $V=0$. 
This seems to suggest that the critical value of $\alpha$ for the breakdown of the full SCBA somehow depends on the 
bias. Moreover at finite bias $\alpha_\mathrm{crit}$ is characterized by a peak in $\rm{\Delta_{thr}}(\alpha)$ for 
positive ${\{r_{\lambda}\}_\mathrm{init}}$ and by a discontinuity for negative ${\{r_{\lambda}\}_\mathrm{init}}$  (see figure \ref{H2FdFull}a).

\begin{figure}[hbt]
\begin{center}
\includegraphics[clip=true,width=.48\textwidth]{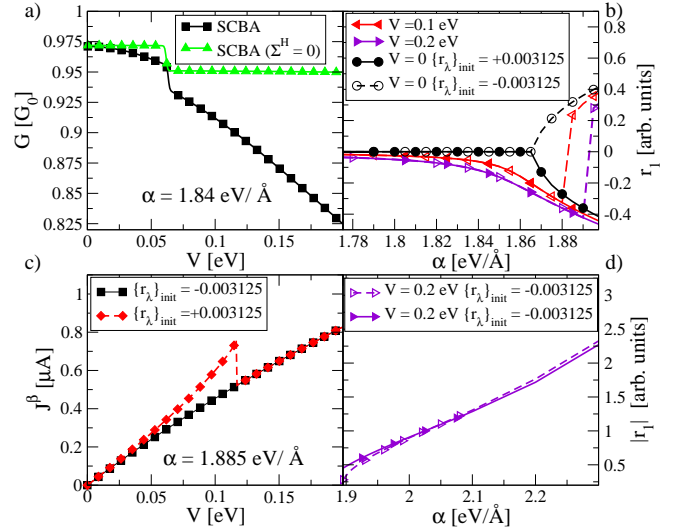}
\end{center}
\caption{\small{Behaviour of the SCBA in the region above and below the zero-bias $\rm{\alpha_{crit}}$. 
For $\rm{\alpha < \alpha_{crit}}$, $G(V)$ for the full and partial SCBA are compared in panel a), showing that the contribution 
arising from $\rm{\Sigma^H}$ is bias-dependent. For $\rm{\alpha > \alpha_{crit}}$, ${J^{\beta}}$ calculated with the full SCBA 
is presented in c) for two different initial conditions. In panel b) the functional dependence of the weighting coefficient ${r_1}$ 
upon bias is investigated for a range of coupling strengths and bias. Finally, in panel d) we show the magnitude ${|r_1|}$ obtained from
two full SCBA simulations with different initial conditions and $\rm{\alpha > \alpha_{crit}}$.}}
\label{biasrcoeff}
\end{figure}

In order to explore the onset of the breakdown of the SCBA at finite bias in figure \ref{biasrcoeff}a we present $G(V)$ for 
${\alpha=1.84}$ eV/\AA\ i.e. just below the zero-bias critical value. In addition, in figure \ref{biasrcoeff}b we plot the 
dominant coefficient ${r_1}$ for a small range of ${\alpha}$ about $\alpha_\mathrm{crit}$ at three different bias, $V=0$, 
$V=0.1$ and $V=0.2$~Volt. A clear result from figure \ref{biasrcoeff}a is that the presence of $\Sigma^\mathrm{H}$ 
introduces a reduction of $G(V)$ with bias not present at $V= 0$. For ${\alpha < \alpha_\mathrm{crit}}$ figure \ref{biasrcoeff}b 
shows that $|r_1|$ is a function of bias whose value increases as the bias increases. The difference between ${r_1}$ at 
$V=0$ and at finite $V$ explains the deviation below $\rm{\alpha_{crit}}$ and also the origin of the peak above it.

The discontinuity in $\rm{\Delta_{thr}}$ for ${\{r_{\lambda}\}_\mathrm{init}>0}$ of figure \ref{H2FdFull}a is explained by the discontinuities 
observed in ${r_1}$ (Fig.~\ref{biasrcoeff}b) and in the current ${J^{\beta}(V)}$ (Fig.~\ref{biasrcoeff}c). In fact at
$V= 0$, $\rm{r_1}$ is single valued as long as $\rm{\alpha < \alpha_{crit}}$. For $\rm{\alpha > \alpha_{crit}}$ instead ${r_1(\alpha)}$ has a parabolic shape 
symmetric about ${r_1=0}$: the ${\{r_{\lambda}\}_\mathrm{init}<0}$ solution traces out the lower arm of the parabola and the 
${\{r_{\lambda}\}_\mathrm{init}>0}$ solution follows the upper arm as $\rm{\alpha}$ increases. For $V>0$, it is seen that the 
two solutions for ${r_1}$ are identical, and asymptotically approach the lower arm of the $V=0$ curve from below for 
${\alpha >\alpha_\mathrm{crit}}$. However, as ${\alpha}$ is further increased the ${\{r_{\lambda}\}_\mathrm{init}>0}$ solution 
jumps discontinuously above zero and then asymptotically approaches the upper arm, again from below.

The discontinuity in ${r_1(\alpha,V)}$ is determined by both the bias and the initial conditions. Generally, it is found that such 
discontinuity occurs for lower bias first; ${r_1}$ jumps discontinuously for $V=0.1$~Volt before it does for $V=0.2$~Volt. 
This explains the behaviour of the ${\{r_{\lambda}\}_\mathrm{init} = +.003125}$ solution for ${J^{\beta}(V)}$ in 
figure \ref{biasrcoeff}c which leads to a peak in its derivative (the conductance $G(V)$) and explains the discontinuity
of $\rm{\Delta_{thr}}$.

We note that after the discontinuity in ${r_1}$ for $V>0$, the two solutions are no longer symmetric about ${r_1=0}$ and
do not converge to the $V=0$ solutions of either arm until ${\alpha\approx 3.0}$~eV/\AA. This is highlighted in 
figure \ref{biasrcoeff}d where ${|r_1|}$ for the two solutions is plotted in a range of ${\alpha}$ just above $\rm{\alpha_{crit}}$ at 
$V= 0.2$~Volt. Such differences explain why $\rm{\Delta_{thr}}$ is not independent of the initial conditions beyond the discontinuity 
and also the different curves observed for $G(V)$ in figure \ref{H2FdFull}c.

We make a final comment about the discontinuities seen in figure \ref{biasrcoeff}a. The value of $\rm{\alpha}$ at which these 
occur is dependent on the initial conditions as mentioned already. Thus for a particular $\rm{\{r_{\lambda}\}_{init}}$ it may 
be possible to reach the upper solution at $\rm{\alpha_{crit}}$ for all bias, so that ${r_1}$ has a parabolic shape for bias while 
being asymmetric about 0. We have not observed this and regard $\rm{\alpha_{crit}}$ as uniquely defined for $V=0$ only.

\section{Results: $\rm{A\lowercase{u}~Chains}$}\label{sec:Auchains}

By using the procedure outlined in section \ref{sec:Num} and the parameters of table \ref{SystemParameters}, the full SCBA 
is used to calculate the transport of the Au RC's. In general we observe a behaviour similiar to that of the H$_2$-Pt system. 
\begin{figure}
\begin{center}
\includegraphics[clip=true,width=.4\textwidth]{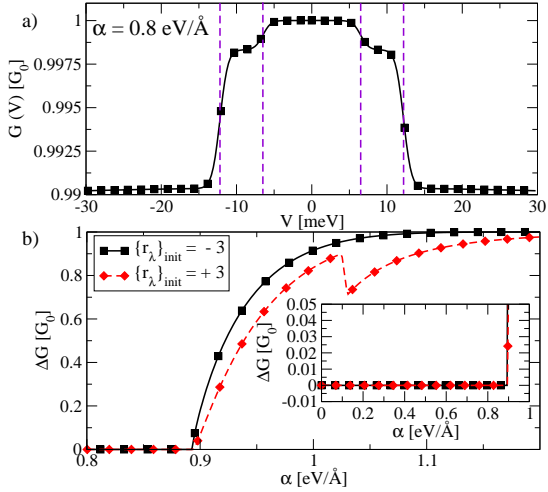}
\end{center}
\caption{\small{$G(V)$ for the 4C illustrating the onset of inelastic processes at threshhold voltages. These are associated
to the symmetric modes of energy $\Omega_2=$~6.5~meV and $\Omega_4=$~12.2~meV. The rigid and 
anti-symmetric modes of energies respectively $\Omega_1=$~3.1~meV and $\Omega_3=$~9.8~meV have no 
effect for chains comprising an even number of atoms. No overall shift in $G(V)$ is observed as $\alpha$ lies below the 
critical coupling $\alpha_\mathrm{crit}$ which is clearly determined from the inset in panel b.}}
\label{diffC.8}
\end{figure}
For 3C, 4C, 5C, and 6C a weak coupling regime is identified where the shift ${\Delta G}$ and the weighting coefficients 
are zero. A critical coupling strength $\rm{\alpha_{crit}}$ for $V=0$ was discovered for each of the chains investigated 
with values $\rm{\alpha_{crit}\approx0.85,0.9,0.82,0.83}$ eV/\AA\ respectively for 3C, 4C, 5C, and 6C. For weak coupling 
${\Delta G}$ matches closely the values calculated in previous works\cite{Freddy1,Freddy2}. As an example, in figure 
\ref{diffC.8} we present $G(V)$ and ${\Delta G (\alpha)}$ for 4C. It is seen that the modes symmetric with respect to the 
centre of the scattering region (even numbered) induce drops in $G(V)$ at threshhold voltages corresponding to the energy 
of the modes. In general, only the symmetric modes are active in RC's containing an even number of atoms and conversely 
the rigid and anti-symmetric modes are active for an odd RC. 
\begin{figure}
\begin{center}
\includegraphics[clip=true,width=.48\textwidth]{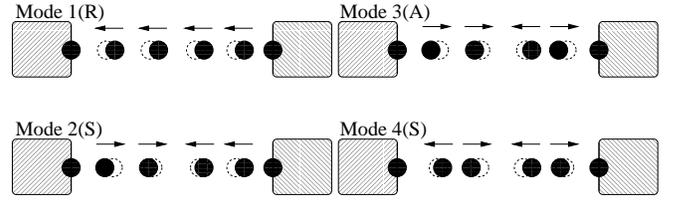}
\end{center}
\caption{\small{Vibration modes of Au 4C. Modes 2 and 4 are symmetric modes (S) about the centre of the 
scattering region. Mode 1 is the rigid translational mode (R) while mode 3 is anti-symmetric (A). For all chains considered, the 
mass of the lead atoms are taken sufficiently larger than that of the chain atoms so that they do not vibrate.}} 
\label{4Cmodes}
\end{figure}

The transition from weak to strong coupling can be appreciated for the 4C by looking at the inset of figure \ref{diffC.8}b,
where for $\alpha>\alpha_\mathrm{crit}$ two different initial conditions lead to two different ${\Delta G}$.
The 6C shows the same behaviour of the 4C, while the 3C and 5C show a single curve for ${\Delta G}$ due to the symmetry 
of the weighting coefficients in the strong regime. Beyond  $\rm{\alpha_{crit}}$, for all the RC's simulated $G(V)$ is reduced 
to zero as ${\alpha}$ increased and the shift ${\Delta G}$ rises to ${G_0}$ .
 \begin{figure}
\begin{center}
\includegraphics[clip=true,width=.48\textwidth]{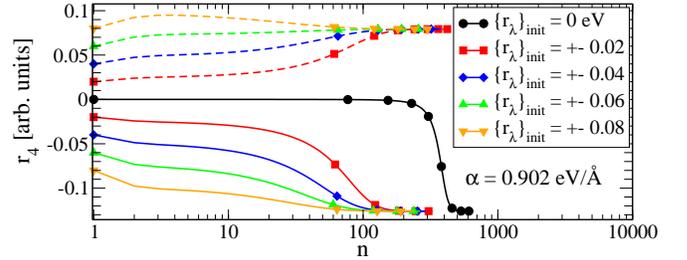}
\end{center}
\caption{\small{Test of the convergence of the weighting coefficient $\rm{r_{4}}$ for the 4C and $\rm{\alpha > \alpha_{crit}}$ using a range $\rm{\{r_{\lambda}\}_{init}}$ where n is the number of iterations. Results for the calculations with positive starting values are indicated with dashed lines while those with negative initial values are shown with solid lines. A single negative and positive solution for  the converged $\rm{r_4}$ is found.}} 
\label{Ic}
\end{figure}
\begin{figure}
\begin{center}
\includegraphics[clip=true,width=.48\textwidth]{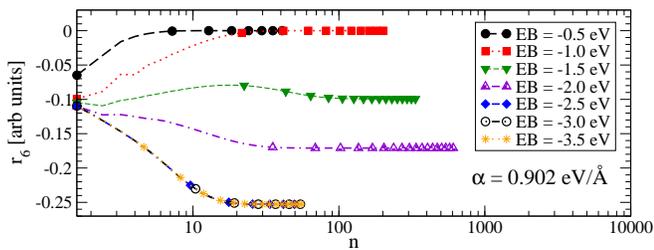}
\end{center}
\caption{\small{Convergence test. The zero-bias ${r_6}$ is plotted versus the iteration number $n$ for 
${\{r_{\lambda}\}_\mathrm{init}= -0.09}$ and ${\alpha > \alpha_\mathrm{crit}}$. Here the lower bound for the
integration of the charge density $\rho$, EB, is varied. The lower band-edge of the leads lies at -2~eV. Clearly
when $\rm{EB \ge -2}$~eV ${r_6}$ does not converge fully, i.e. it depends on EB. }} 
\label{Limitsandgrids}
\end{figure}

Finally we want to investigate further the existence of multiple solutions depending on the initial conditions in the
self-consistent loops. As an example of how convergence is achieved in figure \ref{Ic} we show the coefficient 
${r_4}$ as a function of the iteration number $n$ for the 4C plotted for a single bias $V= 0.02$~Volt and 
coupling 0.902~eV/\AA\ ($\rm{\alpha > \alpha_{crit}}$). A number of simulations were run with different 
initial conditions. The figure indicates the existence of two stable minima: simulations which start at 
${\{r_{\lambda}\}_\mathrm{init}>0}$ (dashed lines) converge to the same positive final value, while simulations 
initialised with ${\{r_{\lambda}\}_\mathrm{init}<0}$ converge to the same $r_6 <0$ value. We note that the minima 
are not symmetric about $r_6=0$. The solution ${r_{\lambda}=0}$ appears to be a minima in the weak coupling 
regime, but becomes unstable and evolves to a local maximum in the strong coupling regime.

In figure \ref{Limitsandgrids} $\rm{r_6}$ is plotted versus the number of iterations $n$ for $\alpha > \alpha_\mathrm{crit}$.
This time we run different simulations in which the lower bound of the energy integration grid, EB, is changed. In particular
we explore situations where EB is not below the lower band-edge of the leads, that for our choice of parameters
lies at -2~eV (see table \ref{SystemParameters}). The figure indicates that if $\rm{EB \ge -1}$~eV ${r_6}$ converges to 
zero so that the sixth mode gives no contribution to $\Sigma^\mathrm{H}$. For $\rm{-2 \le EB < -1}$ ${r_6}$ is nonzero, 
however the converged value differs for EB= -1.5~eV and EB= -2.0~eV, i.e. it is sensitive on the grid lower
bound. Finally for $\rm{EB < -2.0}$ eV the bands of the leads are  entirely included in the integral and ${r_6}$ 
converges to a value of approximately $-0.25$~eV which is independent of our choice of EB. From this simple analysis
it appears that cutting the integration grid can result in the erroneous suppression of the Hartree self-energy, i.e. in a drastic
underestimate of its contribution. This produces a fortuitous suppression of the SCBA breakdown, since the 
agreement between SCBA and EST is usually improved when $\Sigma^\mathrm{H}$ is neglected.

\begin{figure}[htb]
\begin{center}
\includegraphics[clip=true,width=.48\textwidth]{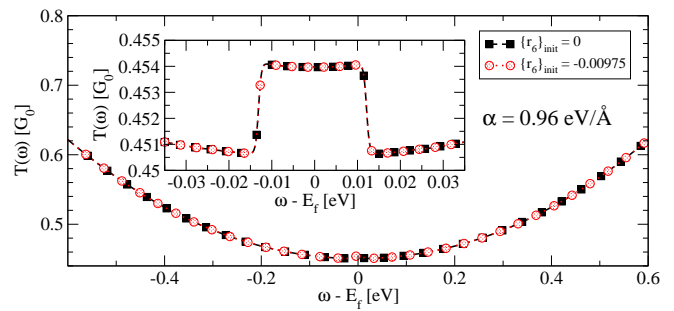}
\end{center}
\caption{$T(\omega)$ for a 6C simulation where only the sixth mode of energy $\rm{\Omega_6=12.6}$~meV is included. 
The curves show results obtained when $\rm{\Sigma^H}$ is calculated using the Simpson's rule 
(${\{r_6\}_\mathrm{init}=0}$ in the legend) or the contour method (${\{r_6\}_\mathrm{init}=-0.00975}$ in the legend). } 
\label{Transm6}
\end{figure}

Finally we test the robustness of our integration method. Figure \ref{Transm6} shows the transmission coefficients 
${T(\omega)}$ for a 6C at a bias of $V=1$~mV where only the sixth mode is considered. 
We run two simulations. In the first case, the full SCBA is used with the numerical parameters taken from table \ref{SystemParameters} and initial condition ${\{r_6\}_\mathrm{init} = -0.00975}$.
In the second case, the integration method outlined in section \ref{subsec:sc} is replaced by Simpson's rule along the real 
axis. The integration range used is [-3.093,0.1]~eV with a grid fineness $\rm{(dE)_{H}=2.129\times 10^{-5}}$~eV. The 
Hartree self-consistent loop is started with ${\{r_6\}_{init} = 0.0}$ and finished when $\rm{c^n_1 < t_H = 10^{-12}}$ 
(this tolerance was also used in the first case). All other parameters of the calculations are identical in the two cases. 
The figure shows that the two numerical methods produce the same ${T(\omega)}$, specifically in the range of 
applied bias shown in the inset.

\section{Conclusions} \label{conclusions}

By using a simple 1D tight binding model we have investigated the breakdown of the SCBA as a function of the
e-p coupling strength $\alpha$. We have identified two regimes. In the weak coupling regime there is a unique solution
for both $\rm{\Sigma^H}$ and $\rm{\Sigma^F}$, independently of the initial conditions. In particular the Hartree self-energy 
is small and has little effect on the final conductance. In this weak coupling regime the characteristic conductance drops
at voltages corresponding to the various phonon energies compare well with those calculated with the EST method.

As the coupling parameter ${\alpha}$ is increased beyond some critical value $\alpha_\mathrm{crit}$ a sharp transition 
to the strong coupling regime occurs. In this limit the self-consistent $\rm{\Sigma^H}$ becomes unstable with respect to 
the initial conditions and exhibits multiple values for the same voltage. This results in a conductance that sharply deviates 
from that obtained with the exact EST. Such a sharp transition suggests a breakdown of the SCBA, indicating that the 
electron-phonon interaction cannot be treated perturbatively for $\alpha>\alpha_\mathrm{crit}$. Interestingly such a
breakdown is suppressed when the Hartree self-energy is neglected completely from the calculation, as routinely done in
practice. Our results however set a warning to quantitative calculations based on e-p parameters extracted from 
density functional theory. For these, no information is available on whether or not the obtained e-p coupling
strength is either below or above the critical value for the SCBA to breakdown. Therefore, strictly speaking,
one rarely knows whether the SCBA is applicable.

\acknowledgments
We acknowledge the financial support of IRCSET and thank Ivan Rungger for all his advice and useful 
discussions. Calculations have been performed at the Trinity Centre for High Performance Computing 
(TCHPC) and at the Irish Centre for High-End Computing (ICHEC).

\end{document}